\documentclass[journal]{IEEEtran}

\usepackage{flushend}
\usepackage{mathrsfs}
\usepackage{color}

\ifCLASSINFOpdf
   \usepackage[pdftex]{graphicx}
\else
  \usepackage[dvips]{graphicx}
\fi

\usepackage{amsmath}

\begin{document}

\title{Experimental Study of a  Planar-integrated Dual-Polarization Balanced SIS Mixer}

\author{Wenlei~Shan,~\IEEEmembership{Member,~IEEE,}
        Shohei~Ezaki,~\IEEEmembership{}
        Keiko~Kaneko,~\IEEEmembership{}
        Akihira~Miyachi,~\IEEEmembership{}
        Takafumi~Kojima,~\IEEEmembership{Member,~IEEE,}
        and~Yoshinori~Uzawa~\IEEEmembership{}
\thanks{Manuscript submitted June 15, 2019.}
\thanks{The work is partly supported by the Japan Society for the Promotion of Science(JSPS) KAKENHI under Grant Number 18K03708}
\thanks{Wenlei Shan, Shohei Ezaki, Keiko Kaneko, Akihira Miyachi, Takafumi Kojima, and Yoshinori Uzawa are with National Astronomical Observatory of Japan, Osawa 2-21-1, Mitaka, 181-8588, Tokyo, Japan (e-mails: wenlei.shan@nao.ac.jp;shohei.ezaki@nao.ac.jp; keiko.kaneko@nao.ac.jp; akihira.miyachi@nao.ac.jp; t.kojima@nao.ac.jp; y.uzawa@nao.ac.jp).}
}

\markboth{\textcopyright 2019 IEEE}%
{Shell \MakeLowercase{\textit{et al.}}: Bare Demo of IEEEtran.cls for IEEE Journals}

\maketitle

\begin{abstract}
A dual-polarization balanced superconductor-insulator-superconductor mixer operating at $2 \,mm$ wavelength is realized in form of a  monolithic planar integrated circuit. Planar orthomode transducers and LO couplers are enabled by using silicon membranes that are locally formed on the silicon-on-insulator substrate. The performance of the balanced mixer is experimentally investigated. Over the entire RF band ($125-163\,GHz$), the balanced mixer shows an LO noise rejection ratio about $15\,dB$, an overall receiver noise about $40\,K$, and a cross-polarization $<-20\,dB$. The  demonstrated compactness and the performance of the integrated circuit indicate that this approach is feasible in developing heterodyne focal plane arrays.
\end{abstract}

\begin{IEEEkeywords}
Radio astronomy, Superconducting integrated circuits, Sensor arrays, Polarimetry.
\end{IEEEkeywords}

\IEEEpeerreviewmaketitle

\section{Introduction}

\IEEEPARstart{C}{oherent} astronomical observations in millimeter (mm) and sub-millimeter (sub-mm) waves using superconductor-insulator-superconductor (SIS) heterodyne receivers enable astronomers to get access to the fine structures of electromagnetic spectra reaching a relative resolution as high as $10^{-7}$ with the highest sensitivity. The spectra contain precise kinematic and chemical information of the celestial objects and the mediums in the transmission pathes. Coherent receiving is also indispensable in interferometry technology to achieve extremely high spatial resolution accompanied with a large collecting area. For example, the event horizon telescope, a very long baseline interferometer working at $1.3\,mm$ wavelength, has reached unprecedented resolution of about $25 \mu \,arcsec $ and led to the first image of the supermassive black hole \cite{EHT2019}. Despite these unique capabilities, the difficulty in constructing focal plane arrays strongly limits the efficiency in the observation of extended sources, particularly for wide field surveys. For example, by using the Delingha $13.7\,m$ telescope, which is equipped with a 9-beam superconducting heterodyne array receiver, it will take about 10 years to finish the Milky Way Imaging Scroll Painting project. This legacy project maps the molecular clouds traced by CO $J=1-0$ transition lines at the Milky Way galactic plane (latitude $< \pm5^o$ and longitude ranging from $-10^o$ to $250^o$)\cite{Su2019},\cite{Shan2012}. To enhance the mapping capability, extensive efforts have been made in the development of heterodyne array receivers for single-dish radio telescopes, about which a comprehensive review can be found in \cite{Groppi2011}. Heterodyne array receivers are also considered to be important in mm and sub-mm interferometers such as ALMA \cite{ALMARoadmap}.

It is a notorious engineering challenge in assembling a compact SIS array receiver by incorporating multiple single-pixel modules as basic building blocks. The difficulty originates from the discrete nature of the modular building blocks, between which there are inevitable interfaces for electrical and mechanical connections. This traditional assembling fashion is not compatible with highly compact integration because the minimum size is limited by the interconnection interfaces. Reliability is also weakened by the complexity in the connections between modules. We have been exploring an alternative approach to the layout for array SIS mixers using monolithic planar integrated circuits, on which silicon membrane-carried probes are locally formed on the chip to couple the signal (in a polarization-sensitive manner) and the local oscillator (LO) \cite{Shan2018a}. The LO coupling approach enables a semi two dimensional metallic waveguide LO distribution network, which can be embedded in the mixer housing structure. In order to demonstrate the feasibility of this method, we have designed, fabricated and evaluated a dual-polarization balanced prototype receiver operating at $2\, mm$ wavelength \cite{Shan2018a},\cite{Shan2018b}. Although the present work involves a single pixel, the design contains all the essential features of the new concept and  the expansion of this work to a multi-pixel array is quite straight forward. In our previous work, we have provided  experimental evidences that indicate the proper operation of the planar orthomode transducer (OMT) on the integrated circuit. The balanced mixing performance, however, was not measured due to unavailability of high quality devices.

In this work, we are able to complete the RF evaluation with devices of improved quality and demonstrate that the planar integrated circuit can provide comparable performance to a traditionally modular type receiver at the measured RF band. In this paper, we begin with a brief revisit of the circuit design. It is followed by the description of the measurement methods. Then the measurement results and related discussion will be given in several aspects, including the receiver sensitivity, the noise rejection ratio (NRR), the balance of the circuit, LO noise measurement, and the crosstalk between polarizations.

\section{Mixer circuit layout and measurement setup}
\begin{figure}[!t]
\centering
\includegraphics[width=3.4in,clip]{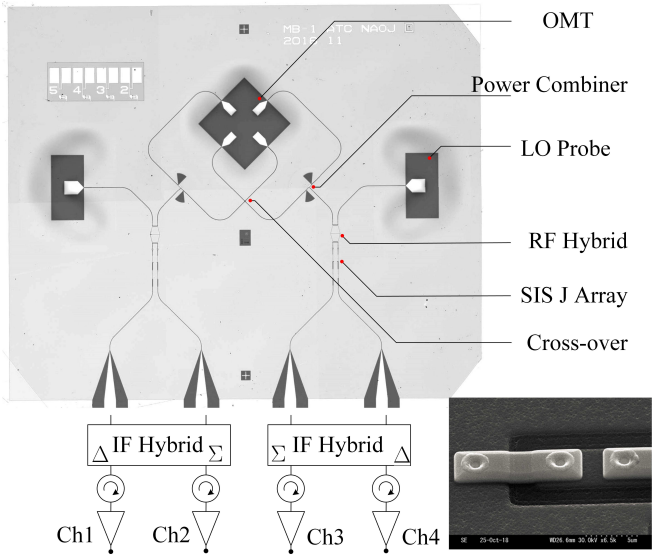}
\caption{The image of the superconducting integrated circuit (chip size $13\,mm\times10\,mm$) together with a diagram of the IF circuit. The inset shows an SEM image of the 3-junction series array (junction size $1.4\mu m$ in diameter). The mixer is designed to enable a direct connection to the LNA. In multipixel applications the isolators can be dropped. See text for details.}
\label{fig1}
\end{figure}

The mixer circuit layout is conceivable from the image of the mixer chip shown in Fig. \ref{fig1}, in which a diagram of intermediate frequency (IF) circuit is also shown to complete the balanced mixing configuration. In the center of the chip  is the planar OMT, which is supported by a locally formed $6\,\mu m$ thick silicon membrane fabricated by removing the handle layer of the silicon-on-insulator (SOI) substrate from the backside. There are two balanced mixers deployed mirror-symmetrically on the chip, each corresponding to one of the two orthogonal linear polarizations. Each balanced mixer comprises a $3 \,dB$ quadrature branch-line coupler and two identical SIS junction arrays each having three junctions connected in series. Inside the LO distribution layer, which is above the chip layer inside the mixer holder, the incoming LO is firstly equally divided for the two polarizations by a waveguide divider, and then they are coupled to the SIS mixers in parallel through the on-chip waveguide probes. Detailed design parameters and theoretical simulation results of these components can be found in \cite{Shan2018b}. The device fabrication is detailed in \cite{Ezaki2018},\cite{Ezaki2019}.

For each balanced mixer, there is a $180^o$ IF hybrid coupler immediately connected to the mixer mount. The IF hybrid couplers used in this study have an amplitude imbalance of $\pm 0.2\, dB$ and a phase imbalanced of $\pm 2^o$ across $4-8\,GHz$ IF band. Normally the two mixers are biased with the same polarity, and consequently the  output signals from the two SIS mixers constructively build up at the $\Delta$ port of the IF hybrid coupler and  cancel out at the $\Sigma$ port. In contrast to the signal, the LO noise appears mainly in the $\Sigma$ port instead of the $\Delta$ port. Each output port of the IF hybrid coupler is followed by a cryogenic isolator-low noise amplifier (LNA) assembly with a passband of $4-8\,GHz$. The LNA has a nominal noise temperature of about $5\,K$.

The mixer is measured in a $4\,K$ cryostat cooled by a GM cryocooler. The standard Y-factor measurement, with two blackbody calibrators at liquid nitrogen temperature and room temperature respectively, is used to measure the receiver noise. The NRR is measured with a CW source, which is weakly coupled into the LO with a $-20\, dB$ waveguide coupler. The cross-polarization of the OMT is measured with a $XY\theta$ beam scanning measurement system outside the cryostat. The measurement results and related discussions will be given in the following sections.

\section{Measurement results and discussion}

\subsection{Receiver sensitivity}
\begin{figure}[!t]
\centering
\includegraphics[width=3.4in,clip]{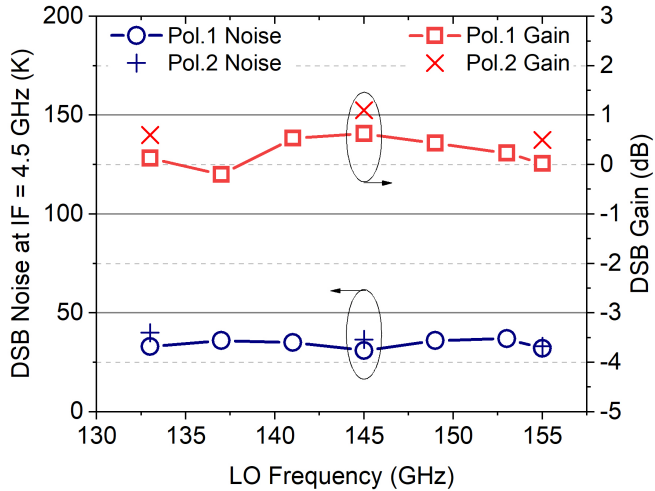}
\caption{The noise temperature (at $IF = 4.5\,GHz$) and mixer gain as a function of the LO frequency. The two polarizations show very similar performance due to the symmetric structure. Data points are only densely measured for polarization No. 1. The RMS error is smaller than the symbol size of the data points.}
\label{fig2}
\end{figure}

For a radio astronomical receiver, its sensitivity is of  primary importance. The measured receiver noise temperature as a function of LO frequency is shown in Fig. \ref{fig2}, together with the mixer conversion gain, which is closely related to the noise. It is noted that the noise temperature presented in this paper is the uncorrected one. The nominal optical loss due to the vacuum window ($25\,\mu m$ Kapton film) and a room-temperature reflective focusing mirror is estimated to be less than $0.1\,dB$. The noise temperature measured at $IF = 4.5\,GHz$ is about $35\,K$ in the RF band ($LO =133-155\,GHz$). The mixer conversion gain is estimated from the ratio of the overall receiver gain to that of the IF chain, assuming a negligible small input loss. The former is measured with a blackbody radiation at room temperature, and the latter is measured by using the tunneling current ($I_t$) as a shot noise calibrator, which has a spectral power density of $2eI_t$  when the SIS junction array is biased at the linear part of the IV curve. The measured receiver noise and mixer conversion gain are favorably flat in the RF band. We attribute this to the relatively high current density (Jc) of the tunneling junction (about $8 kA/cm^2$; as a reference, the Jc of ALMA Band 4 mixer is about $3kA/cm^2$ \cite{Asayama2014},) and the broadband performance of the OMT and the RF coupler. The substantially high current density of the SIS junctions leads to a low Q factor ($Q\approx2$) of the tuning resonators. This Jc value is apparently more than necessary for covering the RF frequency range in the present study. It is chosen because we aim to show the evidence that the planar circuit with high Jc junctions is potentially able to cover much wider RF band although not measured in this study. In fact it is less demanding to achieve broadband performance in the mixer design by using planar type components than metallic waveguide components. This is because, compared with rectangular waveguide modes, quasi-TEM mode of  planar transmission lines have much weaker dispersion and also less influenced by high order propagation modes. In addition, the wide range of the characteristic impedance of the planar transmission lines, ranging from several $\Omega$ of a microstrip line to more than $100\,\Omega$ of a coplanar waveguide, brings considerable flexibility in designing broadband components.

The overall receiver noise of about $35\, K$ is reasonable for a SIS mixer at this frequency range, but it is noticeably higher than the cutting-edge of  low noise advancement, such as about $20\,K$ (DSB) of the ALMA Band 4 receivers\cite{Asayama2014}. The higher noise level is likely due to the relative low conversion gain of the mixers. The embedding impedance of the mixers is intentionally tuned away from the high-gain region in the design, the same method as which was detailed in \cite{Shan2009}. Although empirically a large conversion gain corresponds to low noise temperature, it likely leads to problematic negative dynamic resistance at the bias point, which may cause instability in the operation of the receiver and prohibit the application in astronomical observations, in which receiver stability, instead of the  noise, limits the ultimate sensitivity. Moreover, a large gain also reduces the dynamic range and causes significant mismatch between a mixer and a LNA, which gives rise to undesired ripples in IF spectra.  To ensure a good IF matching, an isolator is commonly inserted between an SIS mixer and a LNA. However, due to their bulky size, the ferrite isolators cannot be utilized in a very compact array. To allow a direct connection between an SIS mixer and a LNA, an output impedance of SIS mixer about $50\,\Omega$ is desirable. This leads to the compromise in the reduction of the mixer gain. Since mixer gain is reversibly correlated to mixer noise, there will be a penalty in the sensitivity if the gain drops. We have measured ALMA Band 4 mixers and verified that the mixer conversion gain is about $3\,dB$ higher than that of the mixer used in this study.

In addition to the direct noise generated in the SIS junctions, the finite transmission loss of the superconducting transmission line between the OMT and the SIS mixer reduces the sensitivity. The loss originates from the thermally activated quasiparticles at a physical temperature of about $4\,K$. The estimated loss is about $0.2\,dB$ for a total electrical length of about $5.5$ wavelengths at $145\, GHz$, which is calculated with SISMA\cite{Shan2018c}, a software for SIS mixer simulation with a implementation of Mattis-Bardeen theory for superconducting transmission line simulation. This loss induces insignificantly $5\%$ increase in the receiver noise. Fortunately, in the potential application of the planar circuit at a higher frequency, this loss will not increase  if the frequency remains below the gap frequency of the superconducting material according to theoretical calculation based on Mattis-Bardeen theory. This is because the loss per wavelength and the overall electric length of the transmission line do not change significantly with frequency. The sub-mm potential is also based on an assumption of low dielectric loss, which strongly depends on material and fabrication process.  At sub-mm regime a dual-polarization planar integrated SIS mixer may even offer better sensitivity than a conventional type, because the adoption of the planar OMT avoids a considerable loss in the waveguide OMT, with a typical value of $0.06 dB/mm$ in the frequency range of $500-700\, GHz$ \cite{Khudchenko2017} at cryogenic temperature.

\begin{figure}[!t]
\centering
\includegraphics[width=3.4in,clip]{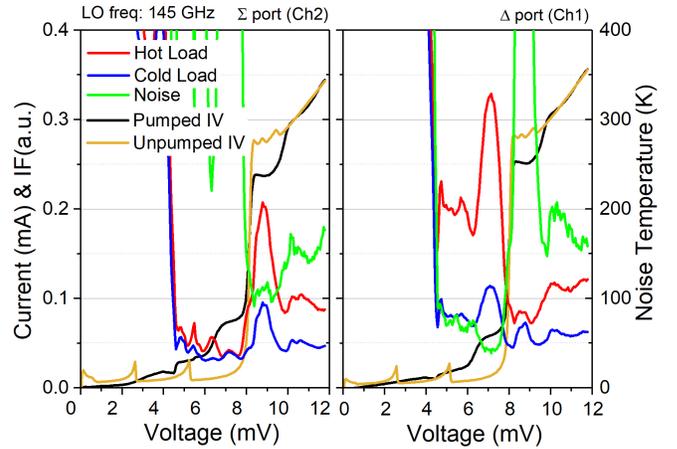}
\caption{The noise temperature, the hot-load and cold-load responses, and the pumped and unpumped IV curves at the $\Delta$ port and the $\Sigma$ port measured at $LO=145\,GHz$. The strong suppression of the signal in the $\Sigma$ port clearly indicates the proper functioning of the balanced mixing.}
\label{fig3}
\end{figure}

The noise temperature and the hot/cold-load responses are recorded at both outputs of the IF hybrid coupler as a function of  bias voltage as shown in Fig. \ref{fig3}. The curves are recorded by sweeping the bias of one SIS mixer in the balanced configuration while fixing the other at the normal bias voltage ($\approx7.2\,mV$). The cancellation of the signal at the $\Sigma$ port is a clear evidence of proper functioning of the balanced mixer. Fixed magnets are used for suppression of Josephson noise. Although a residual DC Josephson current in the umpumped IV curves is observed, it does not affect the noise performance at the normal bias position of the SIS mixer.

\begin{figure}[!t]
\centering
\includegraphics[width=3.4in,clip]{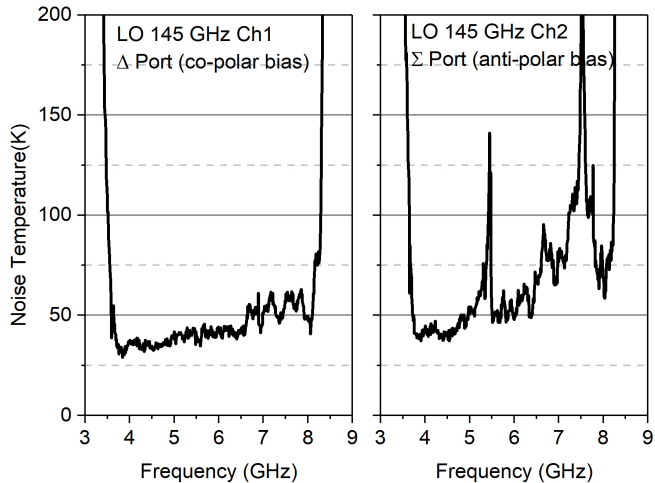}
\caption{The noise temperature measured at $LO=145\,GHz$ as a function of IF. Left panel shows the result measured with co-polar bias of the balanced mixer and the right panel shows the result with anti-polar bias. }
\label{fig4}
\end{figure}

The receiver noise is also measured as a function of IF as shown in Fig. \ref{fig4}. The left panel shows the result with both mixers being biased positively (referred to as co-polar bias), while the right panel shows the result with one mixer being biased positively and the other negatively (anti-polar bias). The difference between the two results will be addressed later in this paper. It should be noted that in the anti-polar bias case the signal turns to appear at the $\Sigma$ port of the IF coupler. Under the normal bias (co-polar), the noise temperature shows a slow increase with IF. This phenomenon is attributed to the capacitance of the DC-blocking capacitor. One terminal of the capacitor connects to the SIS junction lead, and the other terminal connects to the upstream CPW center conductor, which is DC-grounded in the power combiner (T-junction), with detailed layout in \cite{Shan2018b}. As a consequence, the capacitor turns out to shunt the SIS junction array. The undesired narrow IF bandwidth can be corrected by reducing the capacitance without significantly affecting the RF performance.

\subsection{Balance of the circuit, noise rejection ratio and LO noise measurement}

As a major objective of this experimental study, we aim to demonstrate the feasibility of realizing sophisticated SIS mixer circuitry on a single chip, which contains membrane-based probes for LO and signal coupling.  The balanced mixing configuration was chosen because of its favorable feature of saving LO power, which allows pumping a large number of pixels in an array with a single LO source. Another technical reason is that a balance mixer does not require on-chip thin-film resistors, the fabrication process of which has not yet been established in our lab. By contract, for a integrated circuit with practically interesting sideband separation configuration, on-chip loads are necessary to absorb the unused potion of LO power, normally about $90\%$ of the total input.

Noise rejection ratio is the principal figure of merit of a balanced mixer. It is conventionally measured with two individual hot/cold-load measurement procedures, in which the two mixers are applied with co-polar bias and anti-polar bias respectively\cite{Westig2012},\cite{Fujii2017}.  Different bias polarities correspond to constructive or destructive interference of the signals at the output port. This method is based on the underlying assumption that the  bias polarity only affects the relative phase of the two down-converted signals, while keeping the noise and gain of the balanced mixer unaffected. However, in practice, the balanced mixer with anti-polar bias shows noticeably larger ripples in the IF spectrum, as can be seen in the right panel of Fig.\ref{fig4}. This is considered to be due to the undesirable IF impedance levels to the in-phase and out-of-phase IF components from the individual mixers \cite{Kerr2007}. Since the assumption is not fully valid, the accuracy of this method is questionable.

The use of a $180^o$ IF coupler rather than an in-phase power combiner allows us to measure the NRR with the two mixers biased with the same polarity. A weak CW signal is injected into the LO port as an artificial LO noise and down-converted to IF. The mixing outputs are measured at both the $\Delta$ port and the $\Sigma$ port of the IF coupler. The ratio of them is the noise rejection ratio after being corrected for the gain difference of the two IF chains. The gain difference can be measured using the same CW source but with one of the mixer biased above the gap voltage, where the conversion efficiency of that mixer is negligibly small. Under this condition, only one single-ended mixer generates the down-converted CW signal, which is then equally divided by the IF coupler and amplified in the two IF chains. The ratio of the amplitudes of the resulting CW signals, which appear at the output ports of the two IF chains, is the gain ratio to be corrected for.  It deserves a mention that the amplitude of the CW signal is adjusted to be sufficiently weak to avoid saturation of the mixer, but to be relatively strong ($>20\,dB$) with respect to the thermal noise floor, so that the influence of the thermal noise on the measurement is negligibly small.

\begin{figure}[!t]
\centering
\includegraphics[width=3.4in,clip]{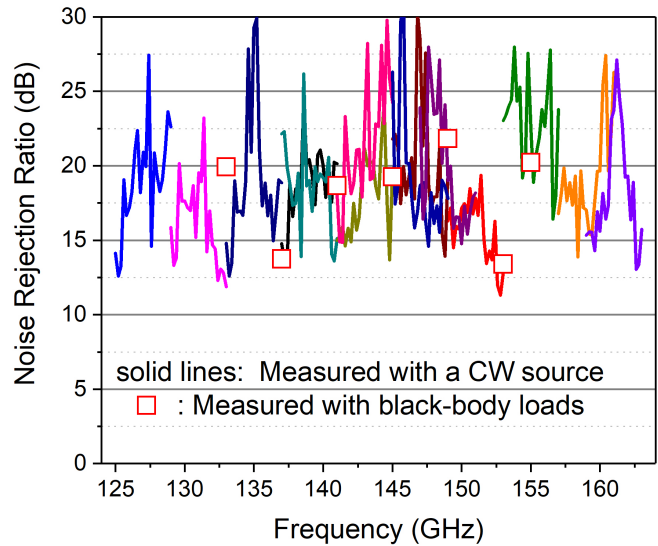}
\caption{The NRR measured with frequency-sweeping of the CW test signal. The results measured with the conventional method are also plotted for comparison. }
\label{fig5}
\end{figure}

The NRR measured with frequency-sweeping of the CW source is plotted in Fig. \ref{fig5}. Unlike the conventional method, where the NRR is an average of both sidebands, the method using CW signal can eliminate the sideband ambiguity. For comparison, the results measured at some discrete frequencies with the conventional method are also shown. The NRR better than $15\, dB$ can be obtained over the RF band with a few exceptions being less than $15\,dB$ but better than $10\,dB$.

 Although the breakdown of the imbalance to each component is difficult due to the tight integration, partial separation is possible. We use the SIS junction arrays as direct detectors to measure the delivered power to each array when a CW source is injected from the signal port or the LO port. The ratio of the responses (photon-assisted tunnelling current) provides the estimation of the imbalance of the RF coupler, supposing the two SIS junction arrays are identical, which is well guaranteed by the high-precision photolithography techniques and verified by comparing their IV curves. Since the ratio, instead of the absolute value, is measured, the CW source does not need to be calibrated for the frequency-dependence of the output power.

\begin{figure}[!t]
\centering
\includegraphics[width=3.4in,clip]{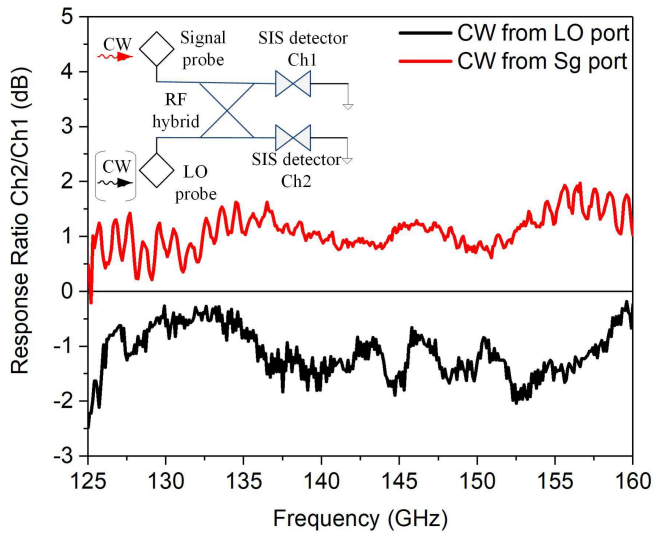}
\caption{The transmission imbalance of the RF coupler measured by direct detection using SIS junction arrays as detectors. A diagram of this measurement configuration is shown in the inset. }
\label{fig6}
\end{figure}

The measured response ratios (Ch2/Ch1) with the CW signal injected respectively from the signal port and the LO port are plotted in Fig. \ref{fig6} as a function of the signal frequency. The mean value is about $\pm1\,dB$ with $1\,dB$ fluctuation over the RF bandwidth. The mean imbalance is probably caused by  inaccuracy in the  design or fabrication errors. This deviation can be compensated for by tuning the insulator thickness of the microstrip lines that compose the RF coupler. According to the theoretical calculation \cite{Kerr2000},\cite{Kooi2012} an imbalance of $1\,dB$ resides in the safe range to achieve a NRR  $>10\,dB$.  Although this method can not provide the phase imbalance, since the phase and the amplitude are not independent, it is justified to believe that the phase imbalance is not significant. The ripples arise due to the mismatching at the input and output ports of the RF hybrid. Because the SIS junction arrays, being weakly pumped and biased below the gap voltage,  must be poorly matched due to their high impedance, the amplitude of the ripples of less than $1\, dB$ indicates reasonably good matching in the input ports.

\begin{figure}[!t]
\centering
\includegraphics[width=3.4in,clip]{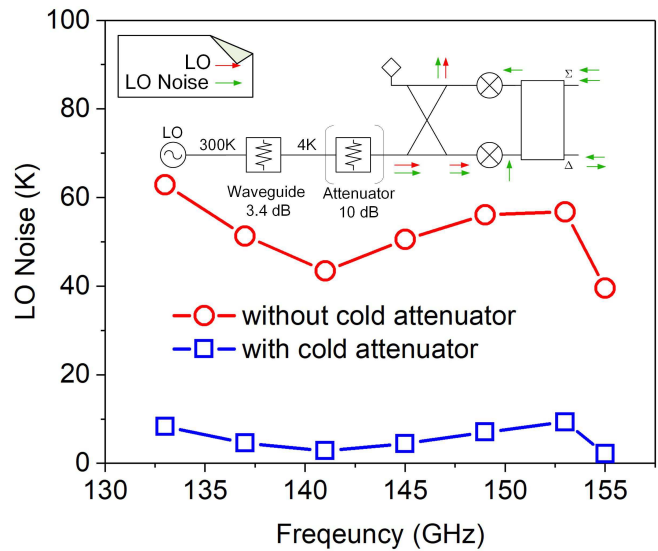}
\caption{The LO noise temperature measured with and without a cold ($4\,K$) $10\,dB$ attenuator. The random measurement error is less than the size of the symbols. The inset shows the diagram of the measurement configuration with small arrows indicating phase relationship.}
\label{fig7}
\end{figure}

The balanced mixer can be used to measure the noise from  LO  and thus is a useful tool in receiver noise diagnosis.  For example, we used this tool to investigate how much thermal noise and non-thermal noise is presented in the LO source used in this study. The LO source is an engineering model of ALMA Band 4 WCA (warm cartridge assembly),  with the final-stage frequency doubler placed at room temperature, rather than placed at $ 110\,K$  stage in an ALMA cartridge receiver. The thermal contribution, an additive noise, can be reduced by a cold attenuator, while the non-thermal noise, which is inherent and proportional to the LO power, cannot. The measurement was done in a procedure similar to the NRR measurement\cite{Westig2012},\cite{Fujii2017}. In order to separate the contributions of the thermal noise and the non-thermal noise, the  measurement is done with and without an additional $10\,dB$ attenuator placed on the $4\,K$ stage immediately before the mixer mount. The measured results are plotted in Fig. \ref{fig7}. It is found that in this specific case almost all of the LO noise is thermal, and the non-thermal portion is zero within the uncertainty of the measurement. The majority of the thermal noise is proved to be dominated by the thermal insulating waveguide, which connects $300\,K$ chamber and the mixer mount at $4\,K$ stage. In the center of of the waveguide is a thermal anchor, which is cooled to a temperature of about $40\,K$. The loss of the waveguide is measured to be about $3.4\,dB$. Supposing the temperature distribution on the waveguide is linear in both halves of the waveguide, we can calculate the input thermal noise with the method explained in the appendix. The result turns out to be that less than $20\,K$ of thermal noise is injected from the LO source outside the cryostat. This reveals that the thermal noise of the LO source is very little.

\subsection{Crosstalk between the two polarizations}

\begin{figure}[!t]
\centering
\includegraphics[width=3.4in,clip]{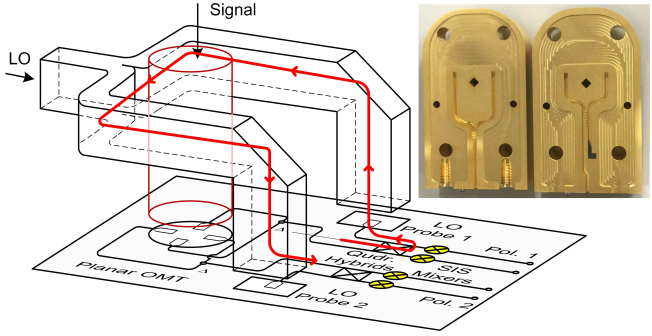}
\caption{The crosstalk between the two polarization channels through the Y-junction waveguide power divider. The replacement of the Y-junction power divider (the left side of the inserted photo) with a branch-line coupler (the right side of the photo) can effectively suppress this crosstalk. }
\label{fig8}
\end{figure}

Stand-alone evaluation of the planar OMT by using a network analyzer is possible at microwave frequencies \cite{Grimes2012}, but practically difficult at mm and sub-mm wavelengthes and at cryogenic temperature. Alternatively we measured the overall cross polarization of the combination of the OMT and a corrugated horn attached to it by using a $XY\theta$ beam scanner. In our previous work, we have demonstrated that a cross polarization at a level $-20\,dB$ is achievable\cite{Shan2018a},\cite{Shan2018b}. However, cross polarization $>-20\,dB$ was observed at some frequencies.  The crosstalk of this level is not likely to arise at the CPW cross-over on the chip (see Fig. \ref{fig1}), because the simulated coupling  $<-30\,dB$ is almost frequency-independent \cite{Shan2018b}. The crosstalk is finally identified to arise through the LO waveguides, as illustrated in Fig. \ref{fig8}. The input signal from one polarization is partly reflected  because of the mismatching from the RF port of the SIS mixer. About half of the reflected signal enters the LO route through the on-chip RF coupler and goes backstream into the LO metallic waveguide through the membrane-based waveguide probe. Because of the insufficient isolation ($-6\,dB$ transmission) between the two output arms of the Y-junction, which divides the LO for the two polarizations, the signal leaks partly from one polarization to the other with a level depending on the input matching condition of the single-ended SIS mixers. To verify this scenario, we replaced the Y-junction power divider with a $3\,dB$ branch-line coupler, which provides $>20\,dB$ isolation. Thanks to this upgrade, the crosstalk is effectively suppressed and the crosstalk level  $<-20\,dB$ can be achieved in the entire RF band.

\section{Conclusion}
With an aim to enhance the capability in forming focal plane arrays with SIS mixers, we have been constructing a frame work of technical solutions to enable planar integration of the SIS receiver frontends. The planar integration concept is implemented in  developing a single chip SIS mixer configured in balanced mixing and dual polarization receiving operating at $2\, mm$ wavelength. The measured receiver sensitivity, the RF bandwidth, and the IF bandwidth are comparable to that of a typical conventional SIS mixer. The LO noise rejection ratio of about $15\,dB$ over the RF band is achievable. The good balance is a favorable consequence of using photolithographic fabrication. The low cross polarization  $<-20\,dB$ in  the entire RF band suggests that the application of planar OMT in mm and sub-mm heterodyne receiver is practical. These results provide clear evidences that the planar integration of SIS mixer focal plane arrays is readily achievable with acceptable compromise between performance and compactness.

\section*{Acknowledgment}
The authors would like to thank Matthias Kroug of NAOJ for his helpful discussion in device fabrication.The authors are also grateful to Daisuke Iono and Alvaro Gonzalez of NAOJ for discussions and advices on potential application in astronomical observations.

\section{Appendix: Thermal noise from a thermal insulating waveguide}
\begin{figure}[!t]
\centering
\includegraphics[width=2.8in,clip]{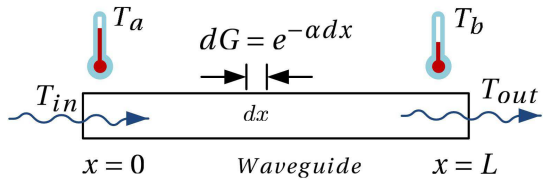}
\caption{calculation model of a lossy waveguide with a linear temperature profile.}
\label{figa1}
\end{figure}

Thermal insulating metallic waveguides made from cupper nickel or stainless steel are often used in cryogenic receiver system for LO injection. The materials with relatively large electric resistance result in substantial loss. In this work, the thermal radiation from the lossy waveguide itself is evaluated using a model shown in Fig. \ref{figa1}.  For a fractional waveguide section with length of $dx$ , the transmission is $e^{-\alpha dx}\approx 1-\alpha dx $ , and the attenuation is $\alpha dx $ , where $\alpha$ is the attenuation constant of the waveguide. According to the radiative transfer equation the differential equation for the radiation is
\begin{equation}\label{a1}
 T(1-\alpha dx)+T_g\alpha dx=T+dT\,,
\end{equation}
where $T$ and $T+dT$ are the input and output radiation equivalent temperature respectively, $T_g$ is the physical temperature of the waveguide section. The waveguide is supposed to connect two stages with temperatures of $T_a$ and $T_b$ respectively, and it is also assumed that the physical temperature is linearly distributed along the waveguide as $T_g= (T_b-T_a)x/L+T_a$, where $L$ is the waveguide length. After normalizing $T$ and $x$ with $T_b-T_a$ and $L$ respectively, the equation becomes,
\begin{equation}\label{a2}
\frac {d\bar T}{d\bar x}=\alpha L(\bar x+\bar T_a-\bar T)\,.
\end{equation}
After introducing a new variable $y$ and defining it as $y\equiv\bar x +\bar T_a-\bar T$, the differential of (\ref{a2}) becomes
\begin{equation}\label{a3}
 \frac{dy}{d\bar x}= 1-\alpha L y\,.
\end{equation}
The integral of (\ref{a3}) with respect to $\bar x$ gives
\begin{equation}\label{a4}
y = \frac{1}{\alpha L}(1-Ce^{-\alpha L \bar x})\,,
\end{equation}
with C being a constant to be determined by the boundary conditions. Combining (\ref{a4}) with the definition of $y$, we finally get
\begin{equation}\label{a5}
\bar T= \bar T_a+\bar x- \frac{1}{\alpha L}(1-Ce^{-\alpha L\bar x})\,.
\end{equation}
Supposing that the radiation is $T_{out} $ at the end point of the waveguide ($\bar x=1 $ ), the constant C can be determined with this condition as
\begin{equation}\label{a6}
C=e^{\alpha L}[1-\alpha L(1+\bar T_a-\bar T_{out})]\,.
\end{equation}
At the input point ($\bar x=0 $ ), the radiation turns out to be
\begin{equation}\label{a7}
\bar T_{in}=\bar T_a-\frac{1}{\alpha L}(1-C)\,.
\end{equation}
The waveguide noise referred to the input port can be calculated as
\begin{equation}\label{a8}
\bar T_{wg}=\frac{\bar T_{out}}{G}-\bar T_{in}=\frac 1G[\bar T_a(1-G)+1+\frac{1-G}{\ln G}]\,,
\end{equation}

where, $G=e^{-\alpha L}$ is the gain of the waveguide.

After this calculation has been done, it is pointed out that the same problem has been studied in \cite{Hu2005}. In that paper, a different method, summation of power series, is used to give the same results. Equation (\ref{a8}) in this work is exactly (7) in \cite{Hu2005}.

\ifCLASSOPTIONcaptionsoff
  \newpage
\fi


\begin{thebibliography}{16}

\bibitem{EHT2019}
The Event Horizon Telescope Collaboration, ¡°First M87 Event Horizon Telescope Results. I. The Shadow of the Supermassive Black Hole,¡± \emph{The Astrophysical Journal Letters}, 875, L1 (17pp), 2019. DOI: 10.3847/2041-8213/ab0ec7
\bibitem{Su2019}
Yang Su, Ji Yang, Shaobo Zhang, Yan Gong, Hongchi Wang, Xin Zhou, Min Wang, Zhiwei Chen, Yan Sun, Xuepeng Chen, ¡°The Milky Way Imaging Scroll Painting (MWISP): Project Details and Initial Results from the Galactic Longitudes of $25^o.8-49^o.7$,¡± \emph{The Astrophysical Journal}, vol. 240, L1(21pp), 2019. DOI: 10.3847/1538-4365/aaf1c8
\bibitem{Shan2012}
Wenlei Shan, Ji Yang, Shengcai Shi, Qijun Yao, Yingxi Zuo, Zhenhui Lin, Shanhuai Chen, Xuguo Zhang, Wenying Duan, Aiqing Cao, Sheng Li, Zhenqiang Li, Jie Liu, Jiaqiang Zhong, ¡°Development of Superconducting Spectroscopic Array Receiver: A Multibeam 2SB SIS Receiver for Millimeter-Wave Radio Astronomy,¡± \emph{IEEE Trans. on Terahertz Sci. and Tech.}, vol.2 , no. 6, pp. 593-604, 2012. DOI: 10.1109/TTHZ.2012.2213818
\bibitem{Groppi2011}
Christopher E. Groppi and Jonathan H. Kawamura, ¡°Coherent Detector Arrays for Terahertz Astrophysics Applications,¡± \emph{IEEE Trans. on Terahertz Sci. and Tech.}, vol. 1, no. 1, pp. 85-96, 2011. DOI: 10.1109/TTHZ.2011.2159555
\bibitem{ALMARoadmap}
¡°The ALMA Development Roadmap,¡± available: https://www.almaobservatory.org/en/publications/the-alma-development-roadmap/
\bibitem{Shan2018a}
Wenlei Shan, Shohei Ezaki, Jie Liu, Shinichiro Asayama, and Takashi Noguchi, ¡°A New Concept for Quasi-Planar Integration of Superconductor-Insulator-Superconductor Array Receiver Front Ends,¡± \emph{IEEE Trans. on Terahertz Sci. and Tech.}, vol. 8, no. 4, pp.472-474, 2018. DOI: 10.1109/TTHZ.2018.2842750
\bibitem{Shan2018b}
W. Shan, S. Ezaki, J. Liu, S. Asayama, T. Noguchi, S. Iguchi, "Planar superconductor-insulator-superconductor mixer array receivers for wide field of view astronomical observation," \emph{Proc. SPIE}, 10708, Millimeter, Submillimeter, and Far-Infrared Detectors and Instrumentation for Astronomy IX, 1070814, 2018. DOI: 10.1117/12.2311933
\bibitem{Ezaki2018}
S. Ezaki, W. Shan, T. Kojima, A. Gonzalez, S. Asayama, and T. Noguchi, "Fabrication and Characterization of Silicon (100) Membranes for a Multi-beam Superconducting Heterodyne Receiver," \emph{Journal of Low Temperature Physics}, vol. 193, pp. 720-725,  2018. DOI: 10.1007/s10909-018-2004-2
\bibitem{Ezaki2019}
S. Ezaki, W. Shan, S. Asayama, and T. Noguchi, "Fabrication of Superconductor Integrated Circuits of D-band Dual-polarization Balanced SIS Mixers," \emph{IEEE Trans. on Appl. Supercond.}, vol. 29 (5), 1101405, 2019. DOI: 10.1109/TASC.2019.2902985
\bibitem{Asayama2014}
Shin¡¯Ichiro Asayama, Toshikazu Takahashi, Kouichi Kubo, Tetsuya Ito, Motoko Inata, Takakiyo Suzuki, Toru Wada, Tomio Soga, Chiyoshi Kamada, Miki Karatsu, Yumi Fujii, Yoshiyuki Obuchi, Susumu Kawashima, Hiroyuki Iwashita, and Yoshinori Uzawa,"Development of ALMA Band 4 (125-163 GHz) receiver," \emph{Publ. Astron. Soc. Jpn} , 66(3), 57(1-13), 2014. DOI: 10.1093/pasj/psu026
\bibitem{Shan2009}
Wenlei Shan, Zhenqiang Li, Jiaqiang Zhong, Shengcai Shi, ¡°An 85-115 GHz SIS Mixer Demonstrating Nearly Constant Dynamic Resistance,¡± \emph{IEEE Trans. on Appl. Supercond.}, vol 19, no. 3, pp. 432-435, 2009. DOI: 10.1109/TASC.2009.2017885
\bibitem{Shan2018c}
Wenlei Shan, Wentao Wu, Shengcai Shi, "SISMA: A Numerical Simulation Software for SIS Mixer Design," \emph{2018 43rd International Conference on Infrared, Millimeter, and Terahertz Waves (IRMMW-THz)}, Nagoya, pp. 1-2, 2018. DOI: 10.1109/IRMMW-THz.2018.8510451
\bibitem{Khudchenko2017}
A. Khudchenko, R. Hesper, A. M. Baryshev, J. Barkhof and F. P. Mena, "Modular 2SB SIS Receiver for 600-720 GHz: Performance and Characterization Methods," \emph{IEEE Trans. on Terahertz Sci. and Tech.}, vol. 7, no. 1, pp. 2-9, Jan. 2017. DOI: 10.1109/TTHZ.2016.2633528
\bibitem{Westig2012}
Marc Peter Westig, Matthias Justen, Karl Jacobs, Jurgen Stutzki, Michael Schultz, Florian Schomacker, and Netty Honingh, ¡°A 490 GHz planar circuit balanced $Nb$-$Al_2O_3$-$Nb$ quasiparticle mixer for radio astronomy: application to quantitative local oscillator noise determination, ¡± \emph{Journal of Applied Physics}, 112, 093919, 2012. DOI: 10.1063/1.4764324
\bibitem{Fujii2017}
Yasunori Fujii, Takafumi Kojima, Alvaro Gonzalez, Shin¡¯ichiro Asayama, Matthias Kroug, Keiko Kaneko, Hideo Ogawa and Yoshinori Uzawa, ¡°Low-noise integrated balanced SIS mixer for 787-950 GHz,¡± \emph{Supercond. Sci. Technol.}, 30, 024001 (13 pp), 2017. DOI: 10.1088/0953-2048/30/2/024001
\bibitem{Kerr2007}
R. Kerr, A. W. Lichtenberger, C. M. Lyons, E. F. Lauria, L. M. Ziurys and M. R. Lambeth, ¡°A superconducting 180 degree IF hybrid for balanced SIS mixers,¡± \emph{Proceedings of the 17th International Symposium on Space Terahertz Technology}, 2007.
\bibitem{Kerr2000}
A. R. Kerr, S.-K. Pan, A. W. Lichtenberger, N. Horner, J. E. Effland, and K. Crady,¡± A single-chip balanced sis mixer for 200-300 GHz,¡± \emph{Proceedings of the 11th International Symposium on Space Terahertz Technology}, pp.251-259, 2000.
\bibitem{Kooi2012}
Jacob W. Kooi, Richard A. Chamberlin, Raquel Monje, Brian Force, David Miller, and Tom G. Phillips, ¡°Balanced Receiver Technology Development for the Caltech Submillimeter Observatory,¡± \emph{IEEE Trans. on Terahertz Sci. and Tech.}, vol. 2, no. 1, pp.71-82, 2012. DOI: 10.1109/TTHZ.2011.2177726
\bibitem{Grimes2012}
P. K. Grimes, O. G. King, G. Yassin and M. E. Jones, "Compact broadband planar orthomode transducer," \emph{Electronics Letters}, vol. 43, no. 21, pp. 1146-1147, 11 Oct. 2007. DOI: 10.1049/el:20071649
\bibitem{Hu2005}
Robert Hu, ¡°Analysis of the input noise contribution in the noise temperature measurements,¡± \emph{IEEE Microwave and Wireless Comp. Lett.} vol. 15, no. 3, pp. 141-143, 2005. DOI: 10.1109/LMWC.2004.842832
\end{thebibliography}
\end{document}